# Full 3D Model of Modulation Efficiency of Complementary Metal Oxide Semiconductor (CMOS) Compatible, Submicron, Interleaved Junction Optical Phase Shifters

Abdurrahman Javid Shaikh[1,*], Fauzi Packeer[2], Mirza Muhammad Ali Baig[1], and Othman Sidek

[1] Department of Electrical Engineering, NED University of Engineering and Technology, Karachi-75290, Pakistan.
[2] School of Electrical and Electronic Engineering, Universiti Sains Malaysia, 14300 Nibong Tebal, Penang, Malaysia.

[*]Corresponding Author
Email: arjs@neduet.edu.pk
ORCID: 0000-0002-8734-3772

**Abstract:** Performance optimization associated with optical modulators requires reasonably accurate predictive models for key figures of merit. Interleaved pn-junction topology offers the maximum mode/junction overlap and is the most efficient modulator in depletion-mode of operation. Due to its structure, the accurate modeling process must be fully three-dimensional, which is a nontrivial computational problem. This paper presents a rigorous 3D model for the modulation efficiency of silicon-on-insulator interleaved junction optical phase modulators with submicron dimensions. Solution of Drift-Diffusion and Poisson's equations were carried out on 3D finite-element-mesh and Maxwell's equations were solved using Finite-Difference-Time-Domain (FDTD) method on 3D Yee-cells. Whole of the modeling process has been detailed and all the coefficients required in the model are presented. Model validation suggests < 10 % RMS error.

## 1. Introduction

Silicon photonics is promising for On-Chip/On-Board (OC/OB) interconnects for its suitable material properties, device and process knowledge, compatibility with mature CMOS processes, possibility of monolithic photonic-electronic integration, and cost effectiveness. Silicon as a material is still under tremendous research for various applications [1,2]. However, due to its centrosymmetric structure, silicon lacks the favourable linear electro-optic (Pockels) effect [3,4]. This leads to free carrier Plasma Dispersion Effect (PDE) as the only viable mechanism for modulation in silicon. Depletion mode silicon optical modulators have been studied extensively in recent years for their high speed and low power consumption as compared to injection mode PIN or accumulation mode Metal-Oxide-Semiconductor (MOS) structures respectively [5,6]. Nevertheless, depletion mode modulators suffer greatly with very small modulation efficiency, resulting in large device footprints – inconsistent with chip-scale photonics. For this, interleaved PN-junction based phase shifter structures were proposed [7]. This topology resulted in modulation efficiency comparable to injection mode modulators, however, compared with horizontal and vertical junction topologies, interleaved junction structure is a compromise between efficiency and bandwidth.

Intricate interdependence and trade-off between several modulator performance parameters exist with respect to the design variables. This necessitates design optimization for a particular application. To define application specific cost function for optimization tasks, performance predictive models are required. There have been attempts in the literature to propose graphical representation of the interleaved phase shifter behaviour, but none provided accurate mathematical predictive models. Moreover, the numerical analysis performed in the literature made non-realistic assumptions which made the analysis interesting for fundamental understanding only. For realistic mathematical models, however, full vectorial and three dimensional (3D) analysis, incorporating all applicable physical models, was required. Such 3D mathematical predictive models would truly enable application specific optimization of interleaved PN-junction optical phase shifters. In this work, we present one of those models, i.e., for modulation efficiency.

## 2. Modelling of Optical Phase Modulators

The perspective view of the device to be modeled is shown in Fig. 1. The optical mode is injected perpendicular to the rib cross section and is propagated along the central elevated rib region where the interleaved PN junction structure is present in the form of alternate P-type and N-type segments or 'fingers'. The injected mode interacts with the local free carrier density and hence phase shift and optical loss occur [4]. The density of free carriers in the path of optical mode is manipulated by the electrical signal applied at the contacts (extreme corners of the structure), modulating phase of the guided mode. The characteristic variation of carrier density in all the three dimensions (including the direction of optical mode propagation) of the interleaved phase modulator calls for the 3D electronic and optoelectronic modeling of the devices. This is a nontrivial problem in photonic device modelling [8-10] and has been thoroughly undertaken for the first time in this work, to the best of our knowledge. Both the symmetric as well as asymmetric PN junctions – involving symmetric and asymmetric doping profiles – have been investigated. All applicable material correction models have been applied during the 3D simulations. All the simulations were performed at 300 K and the wavelength of the mode source was set to 1550 nm.

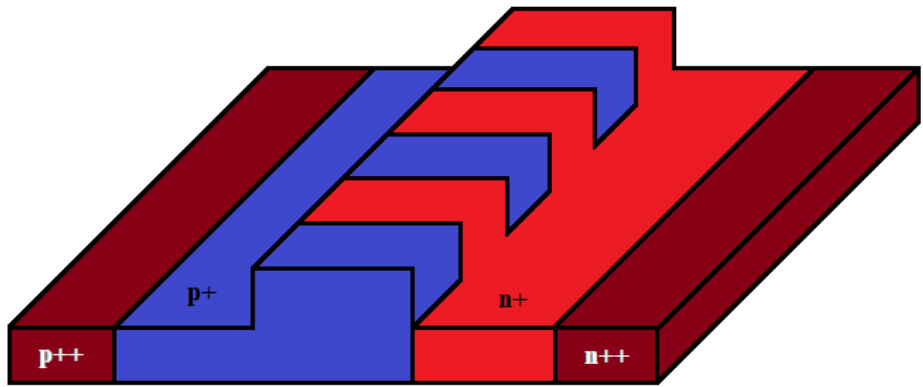

**Fig. 1** Three dimensional perspective view of the interleaved PN junction rib waveguide phase modulator modeled in this work

The modeling process adopted in this paper involves three stages. In the first stage, passive waveguide analysis is performed to identify the single mode boundaries of the submicron Silicon-On-Insulator (SOI) rib waveguide. In the second stage, electronic 3D simulations of the structures have been performed, followed by the third stage which integrates the space dependent charge distribution with the overlap analysis of the injected optical mode to get various performance metrics. For single mode condition and control of minimum confinement factor in submicron SOI rib waveguides, another work of the authors may be referred [11]. The whole modeling flowchart is depicted in Fig. 2.

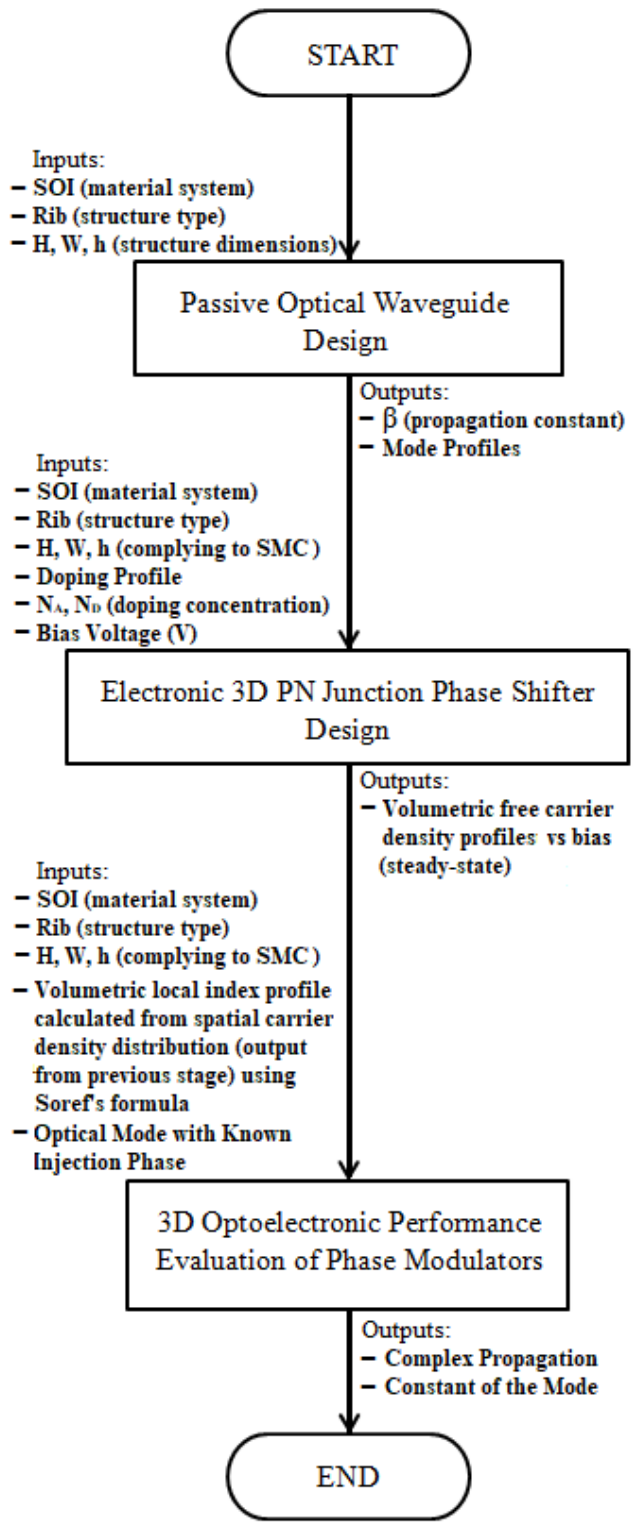

**Fig. 2** Overall analysis workflow of an integrated waveguide phase modulator. The workflow includes three discrete, sequentially interlinked stages

The second stage of the simulation workflow involves analysis of the (active) interaction region (containing P/N-type fingers). Here, and in the third stage, the range on cross sectional dimensions of the waveguide structures to be modeled is determined by the Single Mode Condition (SMC) of the SOI rib waveguide, obtained after completion of the first stage [11]. The goals of the second stage is to generate steady state 3D spatial profiles of free carrier concentration in the waveguide device with respect to bias voltage. For simulations of the device's electrical characteristics, an industry standard finite element mesh based DEVICE TCAD tool has been used [12].

The third stage combines the optical propagation characteristics with the electronic simulation outputs of the modulator device obtained after completion of the second stage. The simulations are 3D as was the case in previous stage simulations. Fundamental mode is injected in the interaction region of the rib waveguide. The goal of this stage is to calculate and record optical mode phase variation as a function of applied reverse bias. This phase variation is then used to calculate the modulation efficiency using following relation.

$$L_\pi = \frac{\pi L}{\Delta \Phi_{SS}} \qquad (1)$$

where, $\Delta \Phi_{SS}$ is the stead state phase shift. This value of $L_\pi$ represents modulation efficiency ($V_\pi L_\pi$) of the phase shifter as the applied voltage signal in this work is restricted to 1 V for CMOS circuit compatibility.

The elaborate orthogonal views and the perspective view of the device to be analyzed are shown in Fig. 3. Fig. 3a shows the top view of the device elaborating the interleaved PN-junction and their respective lengths ($L_P$ and $L_N$). The length '$L_P + L_N$' constitutes the pitch ($\Lambda$) of the modulator. Fig. 3c and 3d show the cross-sectional view from the middle of the P-type finger (the horizontal dotted line marked as '1' in Fig. 3a) and the N-type finger (the horizontal dotted line marked as '2' in Fig. 3a) respectively. The structure between these two cross-section planes shows mirror symmetry in the direction of mode propagation. Therefore, the six surfaces of the cube formed by the volume between the two cross-section planes may serve as the simulation boundary as the DEVICE simulation region conserves charge [12]. This means that simulation of the half pitch length is sufficient (as opposed to full pitch length) which would substantially save computation time and memory requirements.

The structural features enabled by, or defined in terms of, doping; such as junction depth, the finger length in the rib region ($L_P$ and $L_N$) and heavily doped contact distances from the edge of the rib are referred to as *electronic structural parameters* (in this work), as opposed to *waveguide cross-sectional/dimensional/geometric parameters* which are independent of doping [rib height (H), rib width (W), slab height (h)].

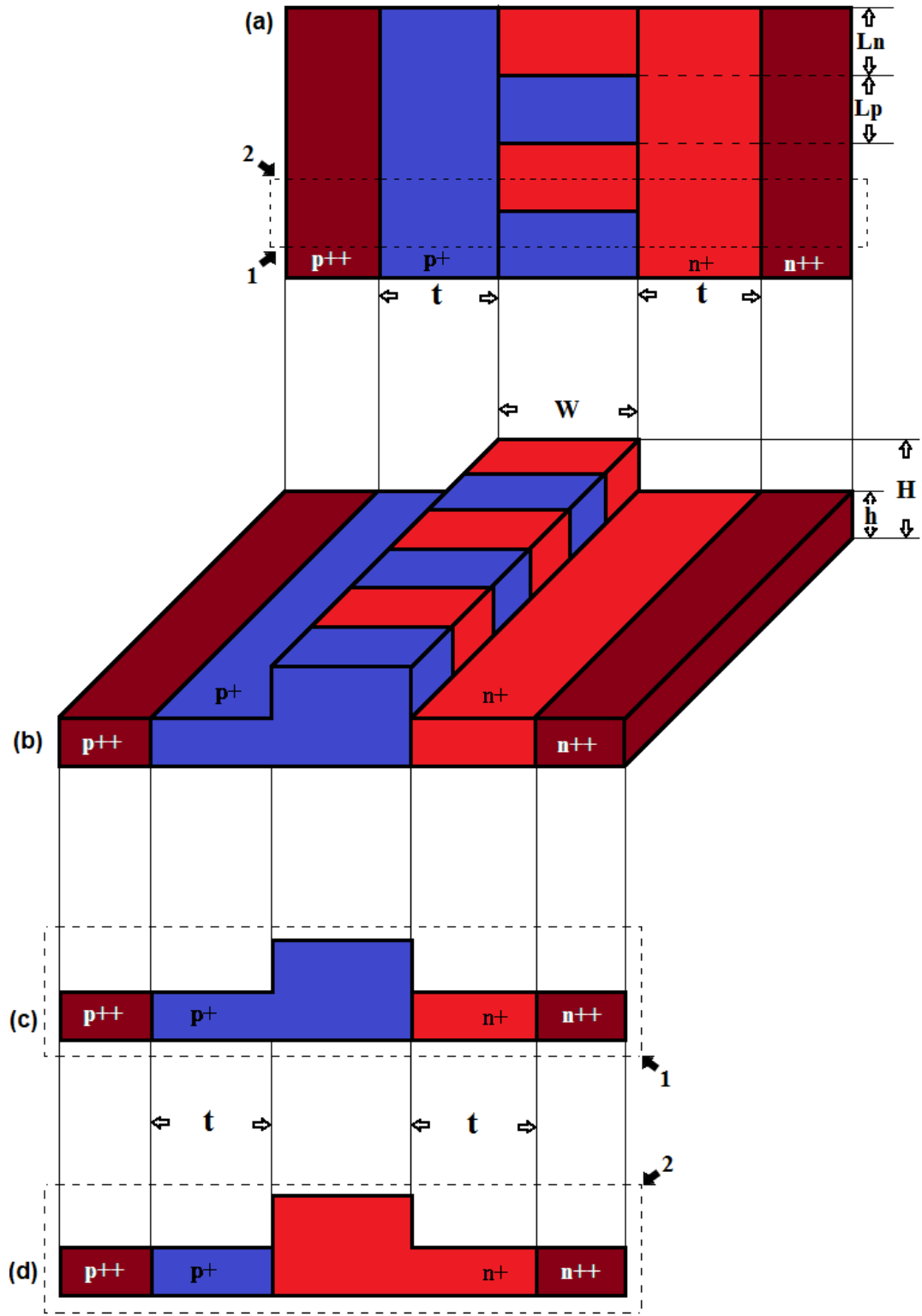

Fig. 3 Orthogonal views (a, c, and d) and the perspective view (b) of the device to be analyzed

## 3. 3D Electronic Simulations

The finite element mesh applied in this work is a non-uniform triangular grid in which the total number of vertices in the structure and the length of each side of a triangle depend on device geometric structure and on the local dopant density. Regions of the device with higher dopant densities require more mesh points to account for the large amount of charge transport at relatively small electric field perturbations. Mesh override regions are deployed in locations where the effect of applied bias may be small but important. Repetitive convergence tests have been performed before finally applying meshing requirements to the device structure.

The dopant density of the N-type finger has been defined in terms of the dopant density of the P-type finger. This has been done to account for both the symmetric ($N_D = N_A$) and asymmetric ($N_D \neq N_A$) junction designs. Moreover, $L_N$ and $L_P$ have been defined in terms of the analytical junction depth ($W_J$) calculated by [13];

$$W_J = \sqrt{\frac{2\varepsilon_0 \varepsilon_r (N_A + N_D)(V_{bi} - V_{app})}{q N_A N_D}} \qquad (2)$$

where, $\varepsilon_0$ is the permittivity of free space, $\varepsilon_r$ is the permittivity of silicon, $V_{bi}$ is the built-in potential, and $V_{app}$ is the applied bias. Selecting finger lengths in this way enabled determination of the effects of working near the fully-depleted junction width operating point [14]. For a fixed waveguide geometry (constant H, W, and h), the number of all the combinations of the electronic structural parameters totals 108 (variable $N_A$, $N_D$, $L_P$, $L_N$, and t – refer Fig. 3.) A total of 24 waveguide geometries have been studied to account for the effects of waveguide rib height, rib width and slab height on the phase shifter efficiency. Hence, the total number of electronic simulations run for the modeling of modulation efficiency amounts to 2592.

After manually defining a particular waveguide geometry, a *core* routine (developed by the authors in the advanced scripting environment) is called which performed electronic structural parameter sweeps over the fixed waveguide geometry. Apart from the electronic parameters sweeps, the routine also adjusted other simulation parameters such as boundaries, simulation mesh and mesh override regions. For each of the 2592 design variations, the steady state spatial charge distributions have been stored in separate files which were imported in the next stage for calculating local refractive index profile for detailed 3D FDTD analysis.

### 3.1 Convergence Testing

Convergence experiment is a cyclic procedure which proceeds by setting moderate values of all the parameters affecting the results, but the one – which needs to be convergence tested for. After getting a converged output for a quantity of interest (or quantities of interest) for the first parameter, the test is run for the second parameter while keeping that value of the first parameter which produced converged results. The process goes on until the quantity(ies) of interest become converged for all the parameters. Then, all the parameters are tested again for convergence starting from the first parameter. If the results for this cycle are the same as were in the first cycle, the cycle is stopped, otherwise a third cycle is run and so on.

Convergence testing of various simulation parameters is must for the results to be meaningful and culminate in a reliable predictive model of the modulator which represents realistic behavior. There are three parameters for steady state simulations against which the convergence tests have been conducted. First, the maximum allowable mesh length in the mesh override region which covers the total analytical space-charge volume. Second, the maximum allowable mesh length in the rest of the simulation region. And, the third, the maximum allowable mesh length in the mesh override regions overlapping the heavily doped contact regions (p++ and n++ regions shown in Fig. 1.) In the steady state simulations, the quantity of interest is the variation in the number of free electrons and holes present in the device before and after applying a small reverse bias dc voltage. If the percentage change in the quantity of interest goes below an already set criterion, then the parameters can be considered producing converged results. In this work, results are considered converged if the percentage error is below 5 %.

## 4. 3D FDTD Optoelectronic Simulations

The 3D optoelectronic simulations, integrating the electronic properties of the device in the optical solver to perform evaluation of the optical figures of merit, has been performed using an industry standard tool based on the *Finite-Difference-Time-Domain (FDTD)* technique [15]. Perfectly Matched Layer (PML) boundary conditions were deployed. The most important step in this stage is to convert the spatial free carrier density calculated in the previous stage to local refractive index distribution. This refractive index distribution is incorporated in the solution of Maxwell's equations through the auxiliary equations relating electric field intensity (**E**) to the electric flux density (**D**). The Free Carrier Index (FCI) and Free Carrier Absorption (FCA) for silicon were first predicted by Soref and Bennett [4,16] which were later updated using the latest experimental data available in 2011 [17]. The FCI and FCA for silicon at 1550 nm are given by [17];

$$\Delta n = -5.4 \times 10^{-22} \Delta N^{1.011} - 1.53 \times 10^{-18} \Delta P^{0.838} \tag{3}$$
$$\Delta \alpha = 8.88 \times 10^{-21} \Delta N^{1.167} + 5.84 \times 10^{-20} \Delta P^{1.109} \tag{4}$$

where, $\Delta N$ is the free-electron density and $\Delta P$ is the free-hole density inside the material. In addition to FCI and FCA, the overall phase shift and the total loss of the propagating mode also depend on the waveguide geometry, therefore, the complex effective index ($N_{eff}$) is the more appropriate quantity which takes into account both the material as well as geometric factors, and is, therefore, calculated using FDTD algorithm.

The *core* routine (developed by the authors in the advanced scripting environment) for the optoelectronic FDTD simulations goes in the similar way as did the core routine for the electronic DEVICE TCAD simulations in the previous stage. The approach in this stage is to import the steady state spatial charge distributions (stored as the output of the previous stage) and convert the spatial charge profile to the spatial refractive index profile. Doing so, now the problem has been converted to that of pure optical one where the electric and magnetic fields are evaluated everywhere in the simulation region by solving Maxwell's equations in time domain on Yee-cells. The 3D mesh in FDTD was dense enough to accommodate most of the vertices generated in the electronic simulations. In this stage, convergence experiments for mesh size, along with other parameters, have been conducted as well.

Two frequency domain field and power monitors are located at the start (along the cross-section plane '1' shown in Fig. 3a) and end (along the cross-section plane '2' shown in Fig. 3a) of the waveguide segment containing the imported charge profile (active segment). The two monitors calculate the transmission characteristics of the mode and field profiles at particular frequency, at their locations. The mode source (at 1550 nm) is located well before the active segment in the passive region of the waveguide. Whole of this arrangement has been depicted in Fig. 4, where the vertical lines '1' and '2' correspond to the cross-section planes labelled as '1' and '2' in Fig. 3a respectively. The steady state charge distribution was imported in the FDTD environment for both the unbiased and reverse biased conditions separately, and the phase variations occurred in the injected mode ($\Delta\phi$) while travelling through the active segment was calculated for both cases. Hence the phase shift achieved after applying the reverse bias is given by;

$$\Delta\Phi_{SS} = \Delta\phi_{0V} - \Delta\phi_{1V} \tag{5}$$

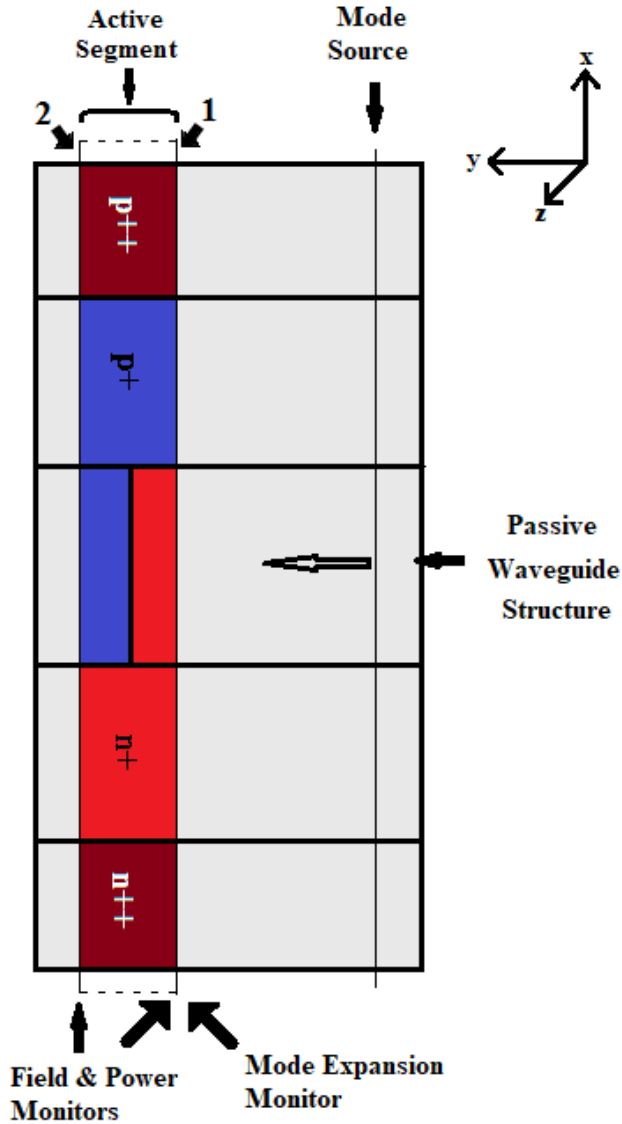

**Fig. 4** Top view of the waveguide modulator structure showing the locations of the active waveguide segment, result monitors and mode source. The light grey/white region is the passive waveguide generated by the FDTD core routine, and the segment between vertical lines '1' and '2' is the electronically active region where the charge distribution had been calculated in stage 2 and converted into the refractive index distribution in this stage

## 5. The predictive model

The predictive model of modulation efficiency has been obtained by systematic stage-wise curve fitting procedures. The modulation efficiency was first fitted with respect to the normalized pitch ($N_\Lambda = \Lambda/W_J$) for constant acceptor and donor dopant densities ($N_A$ and $N_D$). The fitting coefficients obtained therefrom for each of the fixed waveguide geometry and contact-doping/sidewall distance (t) were then fitted with respect to the doping ratio ($r_N = N_A / N_D$) and acceptor dopant concentrations ($N_A$). The modulation efficiency data have been fitted with respect to all the affecting parameters in this way using regression analysis in MATLAB R2015b. The goodness of fit has been taken into account during the fitting procedures for each stage by observing the coefficient of determination ($R^2$) and adjusted coefficient of determination ($\hat{R}^2$). The target values for both the coefficients were more that 90%, however, the fits were much better than this target. No clear trend was observable in the residual plots as well therefore the coefficient values are more reliable.

Based on the careful data generation and goodness of fits, the global model for prediction of the modulation efficiency of CMOS-compatible phase shifters based on interleaved silicon PN-junctions with submicron dimensions is being proposed for the first time as follows;

$$L_\pi(N_A, N_D, L_P, h, W, H) = \frac{N_S N_\Lambda N_0 W_J A_{ME}}{(N_\Lambda + B_{ME})} \sqrt[3]{(1-\Gamma^3)}(1.63 - 1.52H - 0.63W + 0.58HW) \tag{6}$$

where, $N_\Lambda = (L_P + L_N)/W_J$, is the normalized pitch length, $W_J$ is the analytical junction depth, $L_P$ and $L_N$ are P-type finger length and N-type finger length respectively, $N_s$ is the scaling factor and $\Gamma$ is the mode confinement factor. $N_0$ is the factor which takes into account designs with larger junction areas as compared to the one studied in this work, and is given by;

$$N_0(L_S, L_N, L_P, h, W, H) = 1 + \frac{L_S h}{WH + 0.6h\Gamma^{1.5}} \tag{7}$$

where, $L_S$ is the total width of doping region falling outside the edges of the central rib of the waveguide (see Fig. 5).

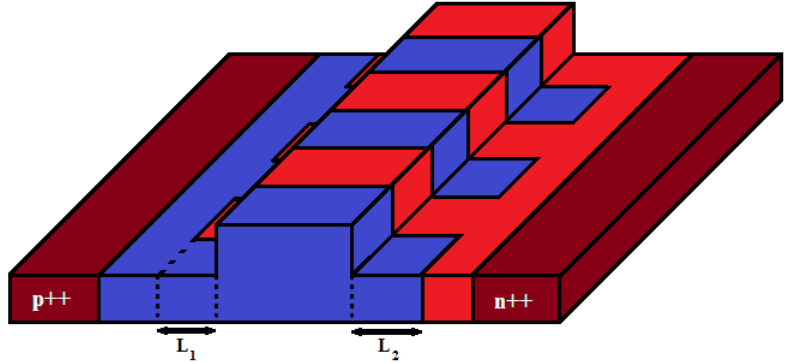

**Fig. 5** Definition of '$L_S$' (refer Eq. 7). $L_S = L_1 + L_2$

The coefficients $A_{ME}$ and $B_{ME}$ in Eq. 6 are the fitting coefficients based on the donor and acceptor dopant concentrations ($N_A$ and $N_D$), and waveguide dimensional parameters ($h$, $W$, and $H$). These coefficients are calculated by following set of equations;

$$\begin{bmatrix} A_{ME} \\ B_{ME} \end{bmatrix} = \begin{bmatrix} a_{00}^{ME} & a_{10}^{ME} & a_{01}^{ME} & a_{20}^{ME} & a_{11}^{ME} & a_{02}^{ME} \\ b_{00}^{ME} & b_{10}^{ME} & b_{01}^{ME} & b_{20}^{ME} & b_{11}^{ME} & b_{02}^{ME} \end{bmatrix} \begin{bmatrix} 1 \\ r_N \\ X \\ r_N^2 \\ r_N X \\ X^2 \end{bmatrix} \tag{8}$$

where $X = log_{10}N_A$, and;

$$a_{00}^{ME}(h, W, H) = \left(p_{00,01_a}^{ME} W + p_{00,02_a}^{ME}\right)h + \left(p_{00,03_a}^{ME} W + p_{00,04_a}^{ME}\right) \tag{9}$$

$$a_{10}^{ME}(h,W,H) = \left(p_{10,01_a}^{ME}W + p_{10,02_a}^{ME}\right)h + \left(p_{10,03_a}^{ME}W + p_{10,04_a}^{ME}\right) \tag{10}$$

$$a_{01}^{ME}(h,W,H) = \left(p_{01,01_a}^{ME}W + p_{01,02_a}^{ME}\right)h + \left(p_{01,03_a}^{ME}W + p_{01,04_a}^{ME}\right) \tag{11}$$

$$a_{20}^{ME}(h,W,H) = \left(p_{20,01_a}^{ME}W + p_{20,02_a}^{ME}\right)h + \left(p_{20,03_a}^{ME}W + p_{20,04_a}^{ME}\right) \tag{12}$$

$$a_{11}^{ME}(h,W,H) = \left(p_{11,01_a}^{ME}W + p_{11,02_a}^{ME}\right)h + \left(p_{11,03_a}^{ME}W + p_{11,04_a}^{ME}\right) \tag{13}$$

$$a_{02}^{ME}(h,W,H) = \left(p_{02,01_a}^{ME}W + p_{02,02_a}^{ME}\right)h + \left(p_{02,03_a}^{ME}W + p_{02,04_a}^{ME}\right) \tag{14}$$

$$b_{00}^{ME}(h,W,H) = \left(p_{00,01_b}^{ME}W + p_{00,02_b}^{ME}\right)h + \left(p_{00,03_b}^{ME}W + p_{00,04_b}^{ME}\right) \tag{15}$$

$$b_{10}^{ME}(h,W,H) = \left(p_{10,01_b}^{ME}W + p_{10,02_b}^{ME}\right)h + \left(p_{10,03_b}^{ME}W + p_{10,04_b}^{ME}\right) \tag{16}$$

$$b_{01}^{ME}(h,W,H) = \left(p_{01,01_b}^{ME}W + p_{01,02_b}^{ME}\right)h + \left(p_{01,03_b}^{ME}W + p_{01,04_b}^{ME}\right) \tag{17}$$

$$b_{20}^{ME}(h,W,H) = \left(p_{20,01_b}^{ME}W + p_{20,02_b}^{ME}\right)h + \left(p_{20,03_b}^{ME}W + p_{20,04_b}^{ME}\right) \tag{18}$$

$$b_{11}^{ME}(h,W,H) = \left(p_{11,01_b}^{ME}W + p_{11,02_b}^{ME}\right)h + \left(p_{11,03_b}^{ME}W + p_{11,04_b}^{ME}\right) \tag{19}$$

$$b_{02}^{ME}(h,W,H) = \left(p_{02,01_b}^{ME}W + p_{02,02_b}^{ME}\right)h + \left(p_{02,03_b}^{ME}W + p_{02,04_b}^{ME}\right) \tag{20}$$

The fitting coefficients $p_{xy,mn_a}^{ME}$, and $p_{xy,mn_b}^{ME}$ in above equations [Eq. (9) – (20)] are the culminating constant values valid for the submicron scale interleaved junction designs based on silicon. For a known confinement factor, these values together with Eqns. (6) – (20) are sufficient to predict the modulation efficiency of the phase shifter. The constant coefficients $p_{xy,mn_a}^{ME}$ and $p_{xy,mn_b}^{ME}$ have been tabulated in Table 1 and Table 2 respectively. The unit of length for the dimensions of rib height (H), rib width (W), slab height (h), etch depth (D), finger length ($L_P$ & $L_N$) is microns, while for doping concentration it is cm$^{-3}$.

Table 1: Model coefficients to calculate the model constant A$_{ME}$.

| $p_{xy,mn_a}^{ME}$ | | xy | | | | | |
|---|---|---|---|---|---|---|---|
| | | 00 | 10 | 01 | 20 | 11 | 02 |
| mn | 01 | -1.330E+05 | 7.535E+03 | 1.430E+04 | 1.843E+02 | -4.621E+02 | -3.807E+02 |
| | 02 | 6.094E+04 | -2.899E+03 | -6.603E+03 | -7.369E+01 | 1.784E+02 | 1.776E+02 |
| | 03 | 1.131E+04 | -8.119E+02 | -1.201E+03 | -1.897E+01 | 4.959E+01 | 3.145E+01 |
| | 04 | -2.709E+03 | 3.518E+02 | 2.737E+02 | 7.595E+00 | -2.135E+01 | -6.664E+00 |

Table 2: Model coefficients to calculate the model constant B$_{ME}$.

| $p_{xy,mn_b}^{ME}$ | | xy | | | | | |
|---|---|---|---|---|---|---|---|
| | | 00 | 10 | 01 | 20 | 11 | 02 |
| mn | 01 | -4.631E+04 | 2.689E+03 | 4.824E+03 | 7.916E+01 | -1.698E+02 | -1.236E+02 |
| | 02 | 1.854E+04 | -1.076E+03 | -1.931E+03 | -3.167E+01 | 6.790E+01 | 4.949E+01 |
| | 03 | 4.773E+03 | -2.771E+02 | -4.971E+02 | -8.151E+00 | 1.749E+01 | 1.274E+01 |
| | 04 | -9.157E+02 | 1.297E+02 | 8.266E+01 | 3.517E+00 | -8.135E+00 | -1.696E+00 |

## 6. Comparison of the model with reported devices

The results obtained from the phase shifter modulation efficiency model developed in section 5 (the model) are compared with the seven designs reported in [18-21]. The particulars of these fabricated designs along with associated confinement factor are tabulated in Table 3. Fig. 6 shows a comparison of the results obtained from the model (solid curve) with the ones reported in various literature (dashed curve.) The overall percentage RMS error of less than 10 % results with the $N_s$ of 0.85 (Eq. 6.)

**Table 3: Several designs fabricated in silicon foundries and reported in various literature.**

| Design No. | Geometric Design Parameters (μm) | | | Electronic Design Parameters | | | | Confinement Factor, Γ (%) | Ref. |
|---|---|---|---|---|---|---|---|---|---|
| | H | W | h | $N_A$ (cm$^{-3}$) | $N_D$ (cm$^{-3}$) | $L_P$ (μm) | $L_N$ (μm) | | |
| 1 | 0.39 | 0.42 | 0.10 | $4\times10^{17}$ | $1\times10^{18}$ | 0.40 | 0.30 | 0.860 | [18] |
| 2 | 0.34 | 0.50 | 0.08 | $2\times10^{17}$ | $2\times10^{17}$ | 0.30 | 0.30 | 0.880 | [19] |
| 3 | 0.22 | 0.45 | 0.15 | $2\times10^{18}$ | $2\times10^{18}$ | 0.30 | 0.30 | 0.550 | [20] |
| 4 | 0.22 | 0.50 | 0.15 | $2\times10^{18}$ | $2\times10^{18}$ | 0.25 | 0.25 | 0.625 | [21] |
| 5 | 0.22 | 0.50 | 0.15 | $2\times10^{18}$ | $2\times10^{18}$ | 0.30 | 0.30 | | |
| 6 | 0.22 | 0.50 | 0.15 | $2\times10^{18}$ | $2\times10^{18}$ | 0.40 | 0.40 | | |
| 7 | 0.22 | 0.50 | 0.15 | $1\times10^{18}$ | $1\times10^{18}$ | 0.30 | 0.30 | | |

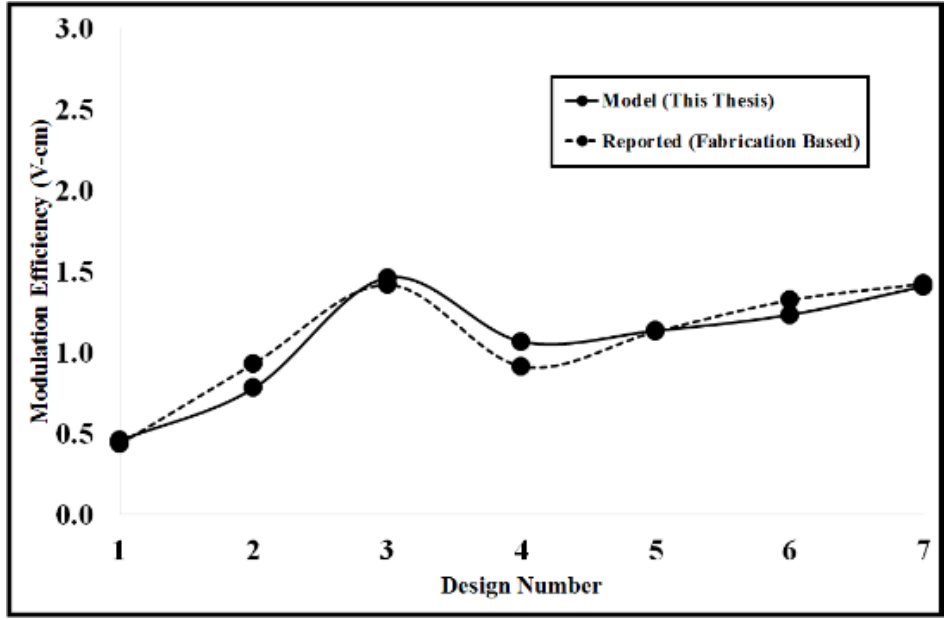

**Fig. 6** A comparison of the results obtained from the model (solid curve) with the ones reported in various literature (dashed curve.) The total RMS error is less than 10%

## 7. Conclusion

Since the interleaved structure's doping and free carrier density profiles are not non-varying with respect to the direction of mode propagation, the problem becomes a three-dimensional one wherein the structure cannot be considered as constant in one of the dimensions. The simulation and modeling performed in this case must be 3D, which is a nontrivial problem in electromagnetics and photonics. We have deployed full 3D optoelectronic simulations and modeling procedure which solved the drift-diffusion, Poisson's and Maxwell's equations in all three spatial dimensions. For solution of Maxwell's equations, 3D finite-difference-time-domain method has been used with perfectly-matched-layer boundary conditions. Maintaining this rigor, we presented a full 3D model for the modulation efficiency of submicron, silicon-on-insulator, interleaved pn-junction optical phase shifters. The bias voltage is limited to a maximum of –1V to maintain CMOS circuit compatibility. The model is strictly for submicron dimensions as it involves polynomials which may manifest Runge's phenomenon for dimensions higher than a micron. All the coefficients required in the model have been presented (Table 2 and 3). The accuracy of the model has then been analyzed by comparing the results obtained by the model with the results reported for the fabricated devices in the literature. An RMS error of less than 10 % suggests the accuracy of the proposed model, given the large number of variables affecting the phase shifter's performance.

**Funding**

This work is partially supported by Universiti Sains Malaysia Research University Short Term Grant (No. 304/PELECT/6315067)